\documentclass[11pt]{article}
\begin{document}
\title{{\bf{\Large Connecting anomaly and tunneling methods for Hawking effect through chirality}}}
\author{
 {\bf {\normalsize Rabin Banerjee}$
$\thanks{E-mail: rabin@bose.res.in}},\, 
 {\bf {\normalsize Bibhas Ranjan Majhi}$
$\thanks{E-mail: bibhas@bose.res.in}}\\
 {\normalsize S.~N.~Bose National Centre for Basic Sciences,}
\\{\normalsize JD Block, Sector III, Salt Lake, Kolkata-700098, India}
\\[0.3cm]
}

\maketitle

\begin{abstract}
   The role of chirality is discussed in unifying the anomaly and the tunneling formalisms for deriving the Hawking effect. Using the chirality condition and starting from the familiar form of the trace anomaly, the chiral (gravitational) anomaly, manifested as a nonconservation of the stress tensor, near the horizon of a black hole, is derived. Solution of this equation yields the stress tensor whose asymptotic infinity limit gives the Hawking flux. Finally, use of the same chirality condition in the tunneling formalism gives the Hawking temperature that is compatible with the flux obtained by anomaly method.
\end{abstract}

{\textbf{\textit {Introduction}}}:
     Ever since Hawking's original observation \cite{Hawking} that black holes radiate, there have been several derivations \cite{Gibbons,Christensen,Paddy,Wilczek,Robinson,Rabin1,Rabin2} of this effect. A common feature in these derivations is the universality of the phenomenon; the Hawking radiation is determined universally by the horizon properties of the black hole leading to the same answer. This, in the absence of direct experimental evidence, definitely reinforces Hawking's original conclusion. Moreover, it strongly suggests that there is some fundamental mechanism which could, in some sense, unify the various approaches.

    In this paper we show that chirality is the common property which connects the tunneling formalism \cite{Paddy,Wilczek,Majhi2,Majhi} and the anomaly method \cite{Christensen,Robinson,Rabin1,Rabin2,Rabin4,Subir,Bonora} in studying Hawking effect. Apart from being among the most widely used approaches, interest in both the anomaly and tunneling methods has been revived recently leading to different variations and refinements in them \cite{Rabin1,Rabin2,Majhi2,Majhi,Rabin4,Subir,Bonora,Morita}. The calculation will be performed using a family of metrics that includes a subset of the stationary, spherically symmetric space-times which are asymptotically flat. Also, the results are derived using mostly physical reasoning and do not require any specific technical skill.

    Before commencing on our analysis we briefly recapitulate the basic tenets of the tunneling and anomaly methods. The idea of a tunneling description, quite akin to what we know in usual quantum mechanics where classically forbidden processes might be allowed through quantum tunneling, dates back to 1976 \cite{Damour}. Present day computations generally follow either the null geodesic method \cite{Wilczek,Majhi2} or the Hamilton-Jacobi method \cite{Paddy,Majhi2,Majhi}, both of which rely on the semiclassical WKB approximation yielding equivalent results. The essential idea is that a particle-antiparticle pair forms close to the event horizon. The ingoing negative energy mode is trapped inside the horizon while the outgoing positive energy mode is observed at infinity as the Hawking flux.

   Although the notion of an anomaly, which represents the breakdown of some classical symmetry upon quantisation, is quite old, its implications for Hawking effect were first studied in \cite{Christensen}. It was based on the conformal (trace) anomaly but the findings were confined only to two dimensions. However it is possible to apply this method to general dimensions. Recently a new method was put forward in \cite{Robinson} where a general (any dimensions) derivation was given. It was based on the well known fact that the effective theory near the event horizon is a two dimensional conformal theory. The ingoing modes are trapped within the horizon and cannot contribute to the effective theory near the horizon. Thus the near horizon theory becomes a two dimensional chiral theory. Such a chiral theory suffers from a general coordinate (diffeomorphism) anomaly manifested by a nonconservation of the stress tensor. Using this gravitational anomaly and a suitable boundary condition the Hawking flux was obtained. A simplified version of this method was given in \cite{Rabin1}. This was followed by another, new, anomaly based approach in \cite{Rabin2,Rabin4}.

    The first step in our procedure is to derive the gravitational anomaly using the notion of chirality. This is a new method of obtaining the gravitational anomaly. Once this anomaly is obtained, the flux is easily deduced. Exploiting the same notion of chirality the probability of the outgoing mode in the tunneling approach will be computed. The Hawking temperature then follows from this probability. At an intermediate stage of this computation we further show that the chiral modes obtained in the tunneling formalism reproduce the gravitational anomaly thereby completing the circle of arguments regarding the connection of the two approaches. 

{\textbf{\textit {Metric and null coordinates}}}:   
   Consider a black hole characterised by a spherically symmetric, static space-time  and asymptotically flat metric of the form,
\begin{eqnarray}
ds^2=F(r)dt^2-\frac{dr^2}{F(r)}-r^2d\Omega^2
\label{1.01}
\end{eqnarray}
whose event horizon $r=r_H$ is defined by $F(r_H)=0$. Now it is well known \cite{Carlip,Robinson} that near the event horizon the effective theory reduces to a two dimensional conformal theory whose metric is given by the ($r-t$) sector of the original metric (\ref{1.01}).

     It is convenient to express (\ref{1.01}) in the null tortoise coordinates which are defined as,
\begin{eqnarray}
u=t-r_*,\,\,\, v=t+r_*;\,\,\,\, dr_*=\frac{dr}{F(r)}.
\label{1.02}
\end{eqnarray}
Under these set of coordinates the relevant ($r-t$)-sector of the metric (\ref{1.01}) takes the form,
\begin{eqnarray}
ds^2=\frac{F(r)}{2}(du~dv+dv~du)
\label{1.04}
\end{eqnarray}
Chiral conditions, to be discussed in the next section, are most appropriately described in these coordinates.

{\textbf{\textit {Chirality conditions}}}:
     Consider the Klein-Gordon equation for a massless scalar particle governed by the metric (\ref{1.04}),
\begin{eqnarray}
g^{\mu\nu}\nabla_\mu\nabla_\nu\phi=\frac{4}{F}\partial_u\partial_v\phi=0.
\label{1.05}
\end{eqnarray}
The general solution of this can be taken as
$\phi(u,v)=\phi^{(R)}(u)+\phi^{(L)}(v)$
where $\phi^{(R)}(u)$ and $\phi^{(L)}(v)$ are the right (outgoing) and left (ingoing) modes satisfying
\begin{eqnarray}
\nabla_v\phi^{(R)}=0,\,\, \nabla_u\phi^{(R)}\neq0;\,\,\,\
\nabla_u\phi^{(L)}=0,\,\,\, \nabla_v\phi^{(L)}\neq0.
\label{1.07}
\end{eqnarray}
These equations are expressed simultaneously as,
\begin{eqnarray} 
\nabla_\mu\phi=\pm\bar{\epsilon}_{\mu\nu}\nabla^\nu\phi
=\pm \sqrt{-g}\epsilon_{\mu\nu}\nabla^\nu\phi
\label{1.08}
\end{eqnarray}
where $+(-)$ stand for left (right) mode and $\epsilon_{\mu\nu}$ is the numerical antisymmetric tensor with $\epsilon_{uv}=\epsilon_{tr}=-1$.
This is the chirality condition {\footnote{In analogy with studies in 2d CFT this condition is usually referred as holomorphy condition.}}. In fact the condition (\ref{1.08}) holds for any chiral vector $J_\mu$ in which case $J_\mu=\pm{\bar{\epsilon}}_{\mu\nu}J^\nu$. Likewise, the chirality condition for the energy-momentum tensor is \cite{Rabin3},
\begin{eqnarray}
T_{\mu\nu}=\pm\frac{1}{2}(\bar{\epsilon}_{\mu\sigma}T^\sigma_\nu+\bar{\epsilon}_{\nu\sigma}T^\sigma_\mu)+\frac{1}{2}g_{\mu\nu}T^\alpha_\alpha
\label{1.09}
\end{eqnarray}
This shows that corresponding to right mode (i.e. ($-$) sign)
\begin{eqnarray}
T^{(R)}_{vv}=0,\,\,\,\,  T^{(R)}_{uu}\neq0
\label{1.10}
\end{eqnarray}   
which is analogous to the first equation of (\ref{1.07}). The other sign of (\ref{1.09}) gives the energy-momentum tensor for the left mode. This essentially corresponds to the interchange $u\leftrightarrow v$  and $L \leftrightarrow R$. In the next section, using these chirality conditions we will derive the explicit form for the gravitational anomaly that reproduces the Hawking flux.     

{\textbf{\textit {Chirality, gravitational anomaly and Hawking flux}}}:
      It is well known that for a non-chiral (vector like) theory it is not possible to simultaneously preserve, at the quantum level, general coordinate invariance as well as conformal invariance. Since the former invariance is fundamental in general relativity, conformal invariance is sacrificed leading to a nonvanishing trace of the stress tensor, called the trace anomaly. Using this trace anomaly and the chirality condition we will derive an expression for the chiral gravitational (diffeomophism) anomaly from which the Hawking flux is computed.

   The energy-momentum tensor near an evaporating black hole is split into a traceful and traceless part by \cite{Fulling1},
\begin{eqnarray}
T_{\mu\nu}=\frac{R}{48\pi}g_{\mu\nu}+\theta_{\mu\nu}
\label{1.11}
\end{eqnarray}
where $\theta_{\mu\nu}$ is symmetric (i.e. $\theta_{\mu\nu}=\theta_{\nu\mu}$), so that it preserves the symmetricity of $T_{\mu\nu}$, and traceless (i.e. $\theta_\mu^\mu=0$ so that in $u,v$ coordinates $\theta_{uv}=0$). The traceful part is contained in the first piece leading to the trace anomaly,
$T^\mu_\mu=\frac{R}{24\pi}$. 
 Also, since general coordinate invariance is preserved, $\nabla^\mu T_{\mu\nu}=0$, from which it follows that the solutions of $\theta_{\mu\nu}$ satisfy,
\begin{eqnarray}
\nabla^\mu\theta_{\mu\nu}=-\frac{1}{48\pi}\nabla_\nu R
\label{1.13}
\end{eqnarray}
   
      Now the energy-momentum tensor (\ref{1.11}) can be regarded as the sum of the contributions from the right and left moving modes. Symmetry principle tells that the contribution from one mode is exactly equal to that from the other mode, only that $u,v$ have to be interchanged. Since $T_{\mu\nu}$ is symmetric we have
$T_{\mu\nu}=T_{\mu\nu}^{(R)}+T_{\mu\nu}^{(L)}$
with
\begin{eqnarray}
T_{\mu\nu}^{(R/L)}=\frac{R}{96\pi}g_{\mu\nu}+\theta_{\mu\nu}^{(R/L)}
\label{1.15}
\end{eqnarray}
where $\theta_{\mu\nu}=\theta_{\mu\nu}^{(R)}+\theta_{\mu\nu}^{(L)}$ (in analogy with $T_{\mu\nu}$). Therefore the chirality condition (\ref{1.10}) and the traceless condition of $\theta_{\mu\nu}$ immediately show
\begin{eqnarray}
&&\theta_{uv}^{(R)}=0,\,\,\, \theta_{vv}^{(R)}=0,\,\,\,\, \theta_{uu}^{(R)}\neq 0;\,\,\, \theta_{uv}^{(L)}=0,\,\,\, \theta_{uu}^{(L)}=0,\,\,\,\, \theta_{vv}^{(L)}\neq 0
\label{1.16}
\end{eqnarray}
The trace anomaly for the chiral components follows from (\ref{1.15}) and (\ref{1.16}),
\begin{eqnarray}
T{^\mu_\mu}{^{(R)}}= T{^\mu_\mu}{^{(L)}}=\frac{1}{2}T^\mu_\mu=\frac{R}{48\pi}
\label{1.17}
\end{eqnarray}

     To find out the diffeomorphism anomaly for the chiral components we will use (\ref{1.15}). Considering only the right mode, for example, we have
\begin{eqnarray}
\nabla^\mu T_{\mu\nu}^{(R)}=\frac{1}{96\pi}\nabla_\nu R+\nabla^\mu\theta_{\mu\nu}^{(R)}
\label{1.18}
\end{eqnarray}
Next, using (\ref{1.13}) and (\ref{1.16}) for the right mode we obtain,
\begin{eqnarray}
\nabla^\mu \theta_{\mu u}^{(R)}=-\frac{1}{48\pi}\nabla_u R;\,\,\, \nabla^\mu\theta_{\mu v}^{(R)}=0
\label{1.19}
\end{eqnarray}
Substituting these in (\ref{1.18}) we get, once for $\nu=u$ and then $\nu=v$,
\begin{eqnarray}
\nabla^\mu T_{\mu u}^{(R)}=-\frac{1}{96\pi}\nabla_u R;\,\,\,\,\, \nabla^\mu T_{\mu v}^{(R)}=\frac{1}{96\pi}\nabla_v R.
\label{1.191}
\end{eqnarray}
Therefore, combining both the above results yields
\begin{eqnarray}
\nabla^{\mu}T_{\mu\nu}^{(R)}=\frac{1}{96\pi}\bar{\epsilon}_{\nu\lambda}\nabla^\lambda R
\label{1.20}
\end{eqnarray}
which is the chiral (gravitational) anomaly for the right mode. Similarly the chiral anomaly for left mode can also be obtained which has a similar form except for a minus sign on the right side of (\ref{1.20}). This anomaly is in covariant form and so it is also called the covariant gravitational anomaly. The structure, including the normalization, agrees with that found by using explicit regularization of the chiral stress tensor \cite{Fulling2,Bat}.

    From (\ref{1.20}) and (\ref{1.17}) a simple relation follows between the gravitational anomaly (${\cal{A}}_\nu$) and the trace anomaly ($T$),
\begin{eqnarray}
{\cal{A}}_\nu=\frac{1}{2}\bar{\epsilon}_{\nu\lambda}\nabla^\lambda T.
\label{grav}
\end{eqnarray}
Such a relation is not totally unexpected since covariant expressions must involve the Ricci scalar. However (\ref{grav}) should not be interpreted as a Wess-Zumino consistency condition which involves only `consistent' expressions \cite{Bat}. Here, on the contrary, we are dealing with covariant expressions.

    The covariant anomaly (\ref{1.20}) is now used to obtain the Hawking flux. As was mentioned earlier the effective two dimensional theory near the horizon becomes chiral. The chiral theory has the anomaly (\ref{1.20}). Taking its $\nu=u$ component we obtain,
\begin{eqnarray}
\partial_rT_{uu}^{(R)}=\frac{F}{96\pi}\partial_r R=\frac{F}{96\pi}\partial_r(F'')=\frac{1}{96\pi}\partial_r(FF''-\frac{F'^2}{2})
\label{new3}
\end{eqnarray}
which yields,
\begin{eqnarray}
T_{uu}^{(R)}=\frac{1}{96\pi}\Big(FF^{''}-\frac{F'^{2}}{2}\Big)+C_{uu}
\label{new1}
\end{eqnarray}
where $C_{uu}$ is an integration constant. Imposing the usual boundary condition which requires $T_{uu}^{(R)}(r\rightarrow r_H)=0$, implying that a freely falling observer sees a finite amount of flux at the outer horizon, leads to $C_{uu}=\frac{F'^2(r_H)}{192\pi}$. This condition on the outgoing modes is similar to that of the Unruh vacuum \cite{Rabin3}. The corresponding condition on the ingoing mode for the Unruh vacuum is satisfied by default since, due to chirality, these are absent ($T_{vv}^{(R)}=0$). Note, however, that the Unruh condition on the ingoing modes $T_{vv}^{(R)}(r\rightarrow\infty)=0$ is applied at asymptotic infinity where the theory is non-chiral. This does not affect our interpretation since, asymptotically, the anomaly (\ref{1.20}) vanishes. Hence the results from the chiral expressions will agree with the non-chiral ones at asymptotic infinity. Indeed, the Hawking flux, obtained by taking the asymptotic infinity limit ($r\rightarrow \infty$) of (\ref{new1}) \cite{Robinson,Rabin1,Rabin2,Rabin4,Bonora},  
\begin{eqnarray}
T_{uu}^{(R)}(r\rightarrow\infty)=C_{uu}=\frac{F'^2(r_H)}{192\pi}=\frac{K^2}{48\pi}
\label{new2}
\end{eqnarray}
where $K=\frac{F'(r_H)}{2}$ is the surface gravity of the black hole, reproduces the known result corresponding to the Hawking temperature $T_H=\frac{K}{2\pi}$. The other terms in (\ref{new1}) drop out due to asymptotic flatness.  

{\textbf{\textit {Chirality, quantum tunneling and Hawking temperature}}}:
    Here, using the chirality condition (\ref{1.08}), we will derive the tunneling probability, which will eventually yield the Hawking temperature. Under the metric (\ref{1.01}) this condition corresponds to,
\begin{eqnarray}
\partial_t\phi(r,t)=\pm F(r)\partial_r\phi(r,t)
\label{1.21}
\end{eqnarray}
As before $+(-)$ stand for left (right) mode. Putting the standard WKB ansatz 
\begin{eqnarray}
\phi(r,t)=e^{-\frac{i}{\hbar}S(r,t)}
\label{1.22}
\end{eqnarray}
in (\ref{1.21}), where $S(r,t)$ is the action, we get the familiar semiclassical Hamilton-Jacobi equation
\begin{eqnarray}
\partial_tS(r,t)=\pm F(r)\partial_rS(r,t)
\label{1.23}
\end{eqnarray}
which is the basic equation in the tunneling mechanism for studying Hawking radiation. This has been derived earlier from the Klein-Gordon equation with the background metric (\ref{1.01}) and the ansatz (\ref{1.22}) \cite{Paddy,Majhi}.

      Now since the metric (\ref{1.01}) is stationary we choose a solution for $S(r,t)$ as
\begin{eqnarray}
S(r,t)=\omega t+S(r)
\label{1.24}
\end{eqnarray}
where $\omega$ is the energy of the particle. Substituting this in (\ref{1.23}) a solution for $S(r)$ is obtained. Inserting this back in (\ref{1.24}) yields,
\begin{eqnarray}
S(r,t)=\omega t\pm\omega\int\frac{dr}{F(r)}
\label{1.25}
\end{eqnarray}
It is important to note that expressing (\ref{1.25}) in the null tortoise coordinates (see (\ref{1.02})) defined inside and outside of the event horizon and then substituting in (\ref{1.22}) one can obtain the right and left modes for both sectors:
\begin{eqnarray}
&&\Big(\phi^{(R)}\Big)_{\textrm{in}}=e^{-\frac{i}{\hbar}\omega u_{\textrm{in}}};\,\,\, \Big(\phi^{(L)}\Big)_{\textrm{in}}=e^{-\frac{i}{\hbar}\omega v_{\textrm{in}}}
\nonumber
\\
&&\Big(\phi^{(R)}\Big)_{\textrm{out}}=e^{-\frac{i}{\hbar}\omega u_{\textrm{out}}};\,\,\, \Big(\phi^{(L)}\Big)_{\textrm{out}}=e^{-\frac{i}{\hbar}\omega v_{\textrm{out}}}
\label{1.251}
\end{eqnarray}
which satisfy the condition (\ref{1.07}). Precisely these modes were  used previously to find the trace anomaly \cite{Fulling1} as well as the chiral (gravitational) anomaly \cite{Fulling2} by the point splitting regularization technique. In our formulation these modes (\ref{1.251}) are a natural consequence of chirality.

     Now in the tunneling formalism a virtual pair of particles is produced in the black hole. One of this pair can quantum mechanically tunnel through the horizon. This particle is observed at infinity while the other goes towards the center of the black hole. While crossing the horizon the nature of the coordinates changes. This can be explained in the following way. The Kruskal time ($T$) and space ($X$) coordinates inside and outside the horizon are defined as,
\begin{eqnarray}
&&T_{\textrm{in}}=e^{K(r_*)_{\textrm{in}}} ~{\textrm{cosh}}(Kt_{\textrm{in}});\,\,\,X_{\textrm{in}}=e^{K(r_*)_{\textrm{in}}} ~{\textrm{sinh}}(Kt_{\textrm{in}})
\nonumber
\\
&&T_{\textrm{out}}=e^{K(r_*)_{\textrm{out}}} ~{\textrm{sinh}}(Kt_{\textrm{out}});\,\,\,X_{\textrm{out}}=e^{K(r_*)_{\textrm{out}}} ~{\textrm{cosh}}(Kt_{\textrm{out}})
\label{Krus1}
\end{eqnarray}  
where, as before, $K=\frac{F'(r_H)}{2}$ is the surface gravity of the black hole. These two sets of coordinates are connected by the relations,
\begin{eqnarray}
t_{\textrm{in}}=t_{\textrm{out}}-i\frac{\pi}{2K};\,\,\,\, (r_*)_{\textrm{in}}=(r_*)_{\textrm{out}}+i\frac{\pi}{2K}
\label{Krus2}
\end{eqnarray}
Therefore following the definition (\ref{1.02}) we obtain the relations connecting the null coordinates defined inside and outside the horizon,
\begin{eqnarray}
&&u_{\textrm{in}}=t_{\textrm{in}}-(r_*)_{\textrm{in}}=u_{\textrm{out}}-i\frac{\pi}{K}
\nonumber
\\
&&v_{\textrm{in}}=t_{\textrm{in}}+(r_*)_{\textrm{in}}=v_{\textrm{out}}
\label{Krus3}
\end{eqnarray}
Since the left moving mode travels towards the center of the black hole, its probability to go inside is 
\begin{eqnarray}
P^{(L)}=|\phi^{(L)}_{\textrm{in}}|^2=1
\label{Krus4}
\end{eqnarray} 
i.e. the left moving (ingoing) mode is trapped inside the black hole.  
On the other hand the right moving mode ($\phi^{(R)}_{\textrm{in}}$) tunnels through the event horizon. To calculate the tunneling probability as seen by an external observer it is first necessary to rewrite $u_{\textrm{in}}$ in terms of $u_{\textrm{out}}$ using(\ref{Krus3}). We find, from (\ref{1.251}),
\begin{eqnarray}
P^{(R)}=|\phi^{(R)}_{\textrm{in}}|^2=|e^{-\frac{i}{\hbar}\omega(u_{\textrm{out}}-i\frac{\pi}{K})}|^2
=e^{-\frac{2\pi\omega}{\hbar K}}
\label{Krus5}
\end{eqnarray}
Then using the principle of ``detailed balance'' \cite{Paddy,Majhi} $P^{(R)}=e^{-\frac{\omega}{T_H}}P^{(L)}=e^{-\frac{\omega}{T_H}}$ yields the Hawking temperature (in the units of $\hbar=1$) as,
\begin{eqnarray}
T_H=\frac{K}{2\pi}
\label{1.30}
\end{eqnarray}
This is the standard Hawking temperature corresponding to the flux (\ref{new2}).

     As we observe the ingoing modes are trapped and do not play any role in the computation of the Hawking temperature. A similar feature occurs in the anomaly approach where the ingoing modes are neglected leading to a chiral theory that eventually yields the flux. These observations provide a physical picture of chirality connecting the tunneling and anomaly methods.      

{\textbf{\textit {Conclusions}}}:
     We have shown that the notion of chirality pervades the anomaly and tunneling formalisms thereby providing a close connection between them. This is true both from a physical as well as algebraic perspective. The chiral restrictions play a pivotal role in the abstraction of the anomaly from which the flux is computed. The same restrictions, in the tunneling formalism, lead to the Hawking temperature corresponding to that flux.

    A dimensional reduction is known to reduce the theory effectively to a two dimensional conformal theory near the event horizon. The ingoing (left moving) modes are lost inside the horizon. They cannot contribute to the near horizon theory thereby rendering it chiral and, hence, anomalous. Using the restrictions imposed by chirality we obtained a form for this (gravitational) anomaly, manifested by a nonconservation of the stress tensor, by starting from the familiar form of the trace anomaly. From a knowledge of the gravitational anomaly we were able to obtain the flux.

   The chirality constraints were then exploited to obtain the equations for the ingoing and outgoing modes in the tunneling formalism, following the standard geometrical (WKB) approximation. We reformulated the tunneling mechanism to highlight the role of coordinate systems in the chiral framework. A specific feature of this reformulation is that explicit treatment of the singularity in (\ref{1.25}) is not required since we do not carry out the integration. Only the modes inside ($\phi_{\textrm{in}}$) and outside ($\phi_{\textrm{out}}$) the horizon, both of which are well defined, are required. The singularity now manifests in the complex transformations (\ref{Krus2}) that connects these modes across the horizon. The probability for finding the ingoing modes was shown to be unity. These modes do not play any role in the tunneling approach which is the exact analogue of omitting them when considering the effective near horizon theory in the anomaly method.

   It is useful to observe that the crucial role of chirality in both approaches is manifested in the near horizon regime. This reaffirms the universality of the Hawking effect being governed by the properties of the event horizon.

\end{document}